# Who Votes for Library Bonds? A Principal Component Exploration.

Eric Jacobson

Weber State University, retired

## ABSTRACT

Previous research has shown a relationship between voter characteristics and voter support for tax bonds.  These findings, however, are difficult to interpret because of the high degree of collinearity across the measures.  From 13 demographic measures of voters in a library bond election, seven independent principal components were extracted which accounted for 95 percent of the variance.  Whereas the direct demographic measures showed inconsistent relationships with voting, the principal components of low SES, college experience, female and service job were related to affirmative voting, while high home value was related to negative voting.

## INTRODUCTION

Special government projects, such as construction of new facilities, are often financed through general obligation bonds.  Such bonds are paid off through direct taxation and generally require the approval of the electorate.  Voter approval indicates that citizens support the project.  Different citizens, though, support different projects.

Beyond knowing that the electorate does, or does not, support the project, it would be helpful to determine the segment of the population that provided the support. This knowledge could lead to a more general understanding of the electorate and its expectations concerning government and could assist local officials plan future development. Several researchers have attempted to identify characteristics of "yes" and "no" voters, but there are serious methodological problems. This paper is an analysis of a bond election to finance the construction of county libraries, an analysis which hopefully avoids some of the previous problems.

This research is exploratory, not confirmatory. No particular view or model of bond elections is tested, and no statistical inferences concerning hypothetical populations of bond elections are possible. Analysis is intended to provide a clear picture of voting behavior in this particular election. Hopefully the results can be a basis for comparison with future studies into bond voting behavior. Extension, modification and perhaps confirmation are to be expected.

## PREVIOUS RESEARCH

*Library Bonds.* Published accounts of library bond elections vary in their objectivity and relevance to voter differences. Some are primarily anecdotal (Charlton 2005; Hall 1990; Lindahl and Berner 1969; Berner 1969). On more solid ground, Adams and Bradley examined the 802 tax related library referenda in Illinois from 1977 to 1995. They noted several relationships between election success and characteristics of the bond proposal, e.g. construction bonds were more likely to pass than other types of tax increments, but did not look at relationships between voter characteristics and voting decisions (Adams and Bradley 1995; Adams 1996). Almost yearly since 1988 Library Journal has published results of a survey of bond elections outcomes, but here too the interest has been solely on effects of the bond proposal and not on individual voter characteristics (Hall 1988; Gold 2004; Freeman 2008; Dempsey 2009; Dempsey 2010; Dempsey 2011; Dempsey 2012; Chrastka and Korman 2015).

Two published studies have explored the relationships between demographic characteristics of voters and support or nonsupport for library funding referenda. Garrison computed correlations between positive votes and demographic properties of voters in voting precincts in two bond elections in Akron, Ohio (Garrison 1963). Cain et al. reviewed all library related ballot measures including bond referenda in California from 1980 to 1995 (Cain et al.1997). They reported correlations and significance levels between total district bond support and voter demographics within that district. The proportion of "yes" votes in these studies was related to the proportion of rental homes and deteriorated houses (Garrison 1963), and to the number of ethnic minority (Cain et al. 1997) within a voting area. On the other hand, support was negatively related to the proportion of voters with service, as opposed to management jobs. Areas with higher incomes (Cain et al. 1997) and higher education levels (Cain et al. 1997; Garrison 1963) gave higher support. Correlations between gender, advanced age and foreign birth, and bond support were low and/or not significant. Reported results were simple correlations, so combined effects of all the variables as might be explored with multiple regression or other techniques are not known.

*School Bonds.* Previous research on library bonds is disappointing both in terms of breadth and depth of statistical analysis. School bond elections, however, have been extensively studied. The set of factors effecting school bonds might well overlap with those effecting library bonds. Nearly 50 years ago Piele and Hall identified 61 factors that might effect school bond referenda (Piele and Hall 1973). Since that time researchers have added several more to the list. This large set of factors can be separated into individual voter factors and referendum election factors. Referendum factors, such as the dollar size of the bond, are constant for a single election and do not concern us here.

Individual voter factors can be divided into two categories: fundamental demographics and societal opinions. Societal opinions, e.g. being cynical toward educators, desiring civic improvement, (Piele and Hall 1973), are problematic. They are inferred from responses to surveys, the reliability and validity of which are always questionable, and since the questions and thus the opinions inferred, differ across studies, results from one bond situation cannot be easily generalized to another. Demographic characteristics, on the other hand, are objective measures which are determined from public records which are independent of the election process. For these reasons we focus on demographic factors and exclude societal opinions.

*Demographic factors.* Since Piele and Hall's review over 20 studies have been published relating demographic characteristics of voters to outcomes on various bond referenda. Some of these examined variation across voting precincts within a single election, while others looked at variation across different districts in different elections. A few used survey techniques in which putative voters were asked how they voted, or would vote, on the ballot issue.

Fourteen demographic factors can be identified in the Piele and Hall review and in subsequent papers. Most of these studies examined more than one factor and most factors have been the subject of several studies. Table 1 lists the 14 factors. (Standardized terms for demographic factors are introduced in this table and are used throughout the paper.) In this research, vote preferences for those with a demographic trait were compared with those of people lacking the trait to determine if it had a positive or negative relationship with bond project approval. Traits which were not naturally dichotomous were made dichotomous with the establishment of threshold criteria, criteria which varied somewhat across research studies. For examples, Low Income was variously defined as "below federal poverty level", or "with children eligible for free school lunch", or "with annual incomes below $25,000," and Minority definitions ranged from the narrow "black or African-American," to the broader "anything but non-Hispanic white." Some studies reported results as cross tabulations or as correlations, but most used some type of multiple regression which measured the combined effects of several factors, usually including factors other than demographic.

**TABLE 1**

**Demographic Measures: Previous Research & Census Sources**

| Measure | Previous research Examples. | Effect | Census Source File | Field | Divisor |
|---|---|---|---|---|---|
| **Married** | Tedin et al. 2001 | ~ | SF1DP1 | HD01_S153 | households |
| **Renter** | Cain et al. 1997 | + | SF1DP1 | HD01_S184 | households |
| **Minority** | Priest and Fox 2005 | + | | sum of below | population |
| *African-Amer* | | | SF1DP1 | HD01_S079 | |
| *Asian* | | | SF1DP1 | HD01_S081 | |
| *Hispanic* | | | SF1DP1 | HD01_S107 | |
| **High Income** | Tosun et al. 2012 | + | DP03 | HC01_VC81-84 | households |
| **Home Value** | Rugh&Trounstine 2011 | ~ | B25077 | All | median |
| **Low Income** | Zimmer et al. 2011 | + | DP03 | HC01_VC75-77 | households |
| **Service Job** | Witt&Pearce 1968 | ~ | DP03 | HC01_VC42 | work force |
| **Manage. Job** | Garrison 1963 | + | DP03 | HC01_VC41 | work force |
| **Unemployed** | Cataldo&Holm 1983 | ~ | DP03 | HC01_VC08 | work force |
| **Female** | Button 1992 | ~ | SF1DP1 | HD01_S051 | population |
| **College** | Plutzer&Berkman 2005 | + | S1501 | Several | population |
| **Parent** | Baldson&Brunner 2003 | + | SF1DP1 | HD01_S152 | households |
| **Old** | Bowers&Lee 2013 | - | SF1DP1 | HD01_S025 | population |
| **Long Term** | Brokaw et al. 1990 | ~ | | | |
| Divisors | | | | | |
| *Population* | | | SF1DP1 | HD01_S001 | |
| *Households* | | | SF1DP1 | HD01_S150 | |
| *labor force* | | | DP03 | HC01_VC05 | |

Notes: Effect in previous research: **+** majority positive effect on voting; **-** majority negative effect; **~** no majority. Measure proportions computed by dividing by indicated divisor.

Not only were a variety of different methods used, but results are not entirely clear or consistent. In some studies, tested factors had no significant effect on voting and effects of some factors have been contradictory, positive in some studies and negative in others. A typical effect for a factor was counted as positive in Table 1 if significant positive outcomes exceeded negative by two, and negative if the reverse. If neither condition held it was entered as neutral.

Characteristics of Renter, Minority, High Income, Low Income, Management Job, College and Parent were related to positive voting. (Support from both High Income and Low Income people suggest a possible U-shaped function with medium income people less supportive.) Only Old was related to negative voting. No other factors had consistent effect.

Table 1 summarizes present knowledge and is a starting point for further research, but it encompasses some statistical shortcomings which could provide a distorted view of the relationship between demographic characteristics and voting. Specifically, in the cited studies problems of predictor collinearity are rarely addressed, and the logical difficulty entailed in the ecological fallacy is not always acknowledged. These issues are addressed in a general statistical context below.

*Digression on statistical analysis.* In most previous research the relationship between vote results and demographics was assessed with some form of regression analysis, modelled as in the following equation

$$\text{bond support} = A + \beta (\text{demo var}) ,$$

With the accuracy of the prediction measured by the squared correlation between the predicted support and the actual support, $R^2$.

The separate relationship between each demographic variable and bond voting can be measured through these "simple" regression analyses, but the separate equations would not indicate the combined effect of all demographic variables. By using sets of independent variables, simple regression is extended to "multiple" regression thusly,

$$\text{bond support} = A + \beta_1 (\text{demo var 1}) + \beta_2 (\text{demo var 2}) + \ldots + \beta_n (\text{demo var n}) \ .$$

With standardized values for the variables, the relative size of the β's indicates the relative strength of the relationship with voting. A demographic variable which has a large effect on voting will have a β with a large absolute value, and a variable with a negligible effect will have a β near zero. Again $R^2$ measures the closeness of the predicted values to the actual.

Demographic variables may be related not just to voting, however, but also to each other. Table 1 shows that both Renter and Low Income have been related to positive voting on bonds. From this we could expect relative high simple regression beta weights for the two variables in the present research. It is plausible, though, that these two variables, Renter and Low Income, are themselves related. The two independent measures may provide redundant information about a voter, so that if they are both included in a multiple regression equation the beta weights would be less than the sum of the two separate simple beta weights. Inter-correlations within a set of independent variables are referred to as multicollinearities.

Without collinearity the $\beta_n$ values in multiple regression will equal β values computed with simple regression. With collinearity, however, the beta weights in multiple regression are not equal to those in simple regression. In fact, beta weights for a given predictor vary depending on which other variables are included in the model. For example, in a study on demographic effects on school bond voting, Strand and associates used Renter and Minority as independent variables (Strand, Giroux, and Thorne 1999). Also as predictors for school bond voting, Bowers and Lee used Minority, Low Income, Old and College (Bowers and Lee 2013). The Strand study found a significant negative effect for Minority, whereas Bowers and Lee found a significant positive effect. These opposite outcomes could be the result of multicollinearity and the differing set of independent variables included in the two analyses. One possible dramatic effect of collinearity is the "suppressor variable," a variable with zero beta when used as the single predictor, but a high beta when included in a multiple regression with other variables (Krus and Wilkinson 1986; Ludlow and Klein 2014)..

When collinearity exists, determining the importance of the relationship between a predictor and a dependent variable is problematic (Farrar and Glauber 1967; Pedhazur 1982). In attempts to overcome the problem, several indices have been developed to estimate predictor importance beyond multiple regression (Nathans, Oswald, and Nimon 2012). One such index is the "relative information weight" (Johnson 2000). A predictor influences the accuracy of the prediction to the extent that it increases $R^2$, an increment which may vary depending on the other predictors included in the regression. The relative information weight of a predictor is the sum of its contribution to $R^2$ in all possible subsets of predictors, expressed as a percentage of the contribution of all predictors across these subsets.

*Principal component regression.* Although work-arounds such as relative information weights have been developed, a more rigorous strategy is to avoid multicollinear predictors in regression altogether. One simple way to accomplish this would be to remove correlated predictors from the model (Pedazur 1982, 246). Another approach is to transform the original set of multicollinear *measured* variables into a set of independent *latent* variables using principal component analysis (PCA) (Jolliffe 1986; Jackson 1991). Similar to factor analysis, the original set of n measures is replaced by a new set of variables defined as weighted sums of the original measures. The procedure derives as many new latent components as original measures and all of the variance in the original measures is retained. The derived components, unlike the original measures, are mutually independent. That is there is no collinearity. It is typical to rotate the axes of the newly derived space in order to clarify the relationship between derived components to measured variables.

PCA derived variables are ordered in terms of how much of the variance in the original measured variables they account for. The first derived component accounts for the most variance, the second component the second most and so on. Derived components farther down on the list may account for very little of the total variance, or information, contained in the measured set of variables and if so, the analysis can be simplified by removing them from further consideration. In principal component regression, vote support is regressed against the set of latent variables, possibly reduced.

In using PCA three issues concerning the latent variables need to be considered. First, how much of the information contained in the measured variables is retained in the latent variable? Those that convey little of the original variance describe little of the demographics of the population of interest. Second, how strong is the relationship of the latent variable with the dependent variable, voting? A latent variable which has a large significant beta in regression may be important even if it accounts for only a small amount of the original variance (Jolliffe 1982). Finally, latent variables are not directly observed, but rather are defined as weighted sums of values on the measured variables. If it is difficult to infer a coherent structure in these additive combinations, then they may provide little insight into the influence of demographic factors.

PCA used in this way is "exploratory." Derived components are descriptions only of the set of data gathered for the particular study. No statistical inferences can be drawn concerning a possible population of similarly gathered data sets. Results are potential guides to analyses of future elections.

*Ecological Fallacy.* To ascertain causal factors in voting the most direct approach would be to record each vote and match it against the demographic characteristics of the voter. The secret ballot makes this impossible. One indirect approach is to ask people how they voted and what their demographic status is. A few researchers have tried this survey method, but it has drawbacks of uncertain reliability and validity. In particular there is evidence that people systematically misrepresent how they vote, either claiming to have voted when they did not, or to have voted in a popular direction when they did not (Clausen 1968; Bernstein, Chadra, and Montjoy 2001; Liu 2007).

An indirect method for *inferring* voter characteristics is to group voters into demographic categories and assess relationships between the group demographic and the group voting proportions. From relationships observed in these averages, relationships for individuals are inferred. If a precinct with a high proportion of Parents also has a high proportion of positive votes, it could be inferred that a Parent is more likely to vote "yes" than a non-Parent. Such an inference goes beyond the data and statistical analysis, and risks serious error, the "ecological fallacy" (Robinson 1950; Jargowsky 2005). Perhaps families with children tend to live in neighborhoods near libraries and people near libraries tend to vote for bonds. Proximity may cause positive voting, not parenthood.

There is no certain logical path by which individual effects can be inferred from averages. Several statistical "solutions" to the ecological fallacy have been proposed (King 1997; Goodman 1959), but they have been applied only to simple regression problems, their effectiveness is disputed and even their authors do not claim them to be infallible (Freedman et al. 1998; King 1997; Jargowski 2005). In this study results from regression and principal component analyses using voters grouped in voting precincts will be reported and then inferences concerning individual behavior will be advanced. The ecological fallacy is risked.

*Analysis Strategy.* The basic unit of measure in this study is the voting precinct. Demographic characteristics for each precinct were obtained from census tables and correlations across these measures were computed. Two regression analyses were then performed. First, vote support was regressed against the measured variables. Ignoring possible collinearity, this analysis produces results which can be directly compared with previous research. Second a principal component analysis was performed and precinct voting outcomes were regressed against the resultant latent variables. Granting plausible interpretations for the latent variables, a clearer relationship between demographics and bond support was revealed.

## METHOD

*Description of Election.* The county where the election took place has an area of 576.1 square miles, a population density of 401.4 people per square mile and a total population of slightly more than 231,000. According to census figures in 2010 the county was 78 percent white and 17 percent Hispanic. 23.2 percent had a Bachelor's degree or higher. 7.4 percent were unemployed and 12.3 percent were below the poverty level. The median age was 30.6; the median income was $56,216; and the median home value was $169,200. The county includes a small central city, suburban communities, agricultural areas, and undeveloped mountain terrain.

In 2013 county commissioners proposed a major building program to replace and/or enhance three overcrowded and deteriorating branch libraries. The project was to be financed by a $45 million bond issue which would increase annual taxes on a median home by $31.50. A special election was set for June of 2013. For the six other summer elections held from 2007 through 2013 the median voter turnout was 13 percent with a maximum in June of 2012 of 20 percent. For the first time in the county, mail voting was used.

County commissioners, library board and staff and an independent group, Friends of the Library, mounted an education campaign to inform voters of the intended improvements and consequent tax implications. This campaign included public meetings; presentations to city councils, Chambers of Commerce, service clubs, and PTAs; newspaper ads; and public information booths. No organized opposition to the bond was in evidence, although the major county newspaper editorialized against it.

Election turnout was extremely high with 30 percent of registered voters voting compared with the median participation of 13 percent over the previous summer elections. The bond issue passed with 17,090 "yes" votes and 14,296 "no."

*Voting and Demographics.* Of the 162 county voting precincts, five had fewer than 10 registered voters and were removed from further analysis. In the retained 157 precincts the mean number of registered voters was 668 and the median 710.

Demographic information was obtained from the Census Bureau in four American Community Survey tables, dated 2012: 2010 Census--SF1DP-1, Education--S1501, Employment--DP03 and Housing--B25077; (download from www.census.gov, February and June, 2014). Length of time in present residence was not available, so analysis was restricted to 13 of the 14 original demographic variables.

Table 1 lists the demographic factors and the census measure used to represent them. Some variables were the sum of several census fields. For example, people with College is the sum of census categories for some college, AA degree, bachelor degree and advanced degree for both 18 to 24 year olds and people older than 24. Counts for each of the 13 variables and for the total number of people, households and people in the work force were obtained for each census tract.

Unfortunately, area boundaries for voting precincts do not correspond with the boundaries for census tracts; a given precinct may be included in several tracts and vice versa. To solve this problem of zonal mismatch, the method of areal weighting was used to estimate the demographic characteristics of the precincts. See Flowerdew and Green for this and other similar methods (1994).

Geographic Information System (GIS) files with voting precincts boundaries were obtained from the state GIS service (downloaded from gis.utah.gov, July 2014). Similarly, GIS Tiger Line files for census tract boundaries were obtained from the Census Bureau (downloaded from www.census.gov, July 2014). Using the GIS system ARC/INFO, the intersections of the two sets of boundaries were plotted, thereby creating all of the subsections contained within both sets of boundaries. Areas within the resulting 750 subsections were then computed.

Many of the subsections were very small, most likely due to disparate data entries for common boundaries in the state and census geography files. In order to avoid confusion from these discrepancies, all subsections of less than 100 square meters were removed from the analysis. The 404 segments remaining after this removal still covered more than 99.99 percent of county area.

Subsections represent the areas within each census tract assigned to a different voting precinct. The proportion of each tract's area included in its component precinct was computed and the census numbers associated with the tract were allocated to each subsection equal to these proportions. Census numbers were then combined across the subsections within each voting precinct to estimate demographic counts and means for voting precincts. Proportions were computed using appropriate census totals: persons, households, and people in the work force. The Minority factor was represented by a sum of Blacks, Asians and Hispanics. In order to better meet distributional requirements for inferential statistics, log values of proportions were used.

Four analyses were performed: (1) pairwise correlations were computed across all demographic variables; (2) voting proportions were regressed against each variable alone and all variables together, and relative information weights were computed; (3) principal components were derived from the demographic variable correlation matrix; (4) voting proportions were regressed against them. Computations were done with SPSS version 13, enhanced with a routine for computing relative information weights (Lorenzo-Seva, Ferrando, and Chico 2010).

## RESULTS

Predictor inter-correlations are shown in Table 2. Collinearity abounds. Almost all variables are correlated with almost all other variables. Some of these correlations are very high and most are significant at the .05 level. The single exception is Female which seems to be independent of most other variables.

**TABLE 2**

**Correlations between Demographic Measures**

|  | Mar | Ren | Min | HiI | HmV | LoI | Ser | Man | Unem | Fem | Col | Par | Old |
|---|---|---|---|---|---|---|---|---|---|---|---|---|---|
| Married |  | **-.90** | **-.84** | **.93** | **.74** | **-.84** | **-.68** | **.76** | **-.66** | -.12 | **.47** | -.12 | .03 |
| Renter | **-.90** |  | **.77** | **-.81** | **-.57** | **.85** | **.66** | **-.62** | **.56** | .14 | **-.30** | .01 | .06 |
| Minority | **-.84** | **.77** |  | **-.88** | **-.79** | **.62** | **.52** | **-.84** | **.64** | -.02 | **-.63** | **.43** | **-.35** |
| High Inc. | **.93** | **-.81** | **-.88** |  | **.82** | **-.78** | **-.60** | **.87** | **-.67** | -.10 | **.62** | **-.28** | .18 |
| Home Val. | **.74** | **-.57** | **-.79** | **.82** |  | **-.55** | **-.41** | **.81** | **-.58** | -.09 | **.62** | **-.36** | .20 |
| Low Inc. | **-.84** | **.85** | **.62** | **-.78** | **-.55** |  | **.65** | **-.55** | **.51** | .09 | **-.30** | -.11 | .18 |
| Serv. Job | **-.68** | **.66** | **.52** | **-.60** | **-.41** | **.65** |  | **-.50** | .27 | .04 | -.16 | -.19 | .17 |
| Mnge. Job | **.76** | **-.62** | **-.84** | **.87** | **.81** | **-.55** | **-.50** |  | **-.59** | .01 | **.79** | **-.50** | **.41** |
| Unempl. | **-.66** | **.56** | **.64** | **-.67** | **-.58** | **.51** | .27 | **-.59** |  | .13 | **-.45** | **.19** | -.06 |
| Female | -.12 | **.14** | -.02 | -.10 | -.09 | .09 | .04 | .01 | .13 |  | .07 | **-.25** | **.30** |
| College | **.47** | **-.30** | **-.63** | **.62** | **.62** | **-.30** | -.16 | **.79** | **-.45** | .07 |  | **-.70** | **.59** |
| Parent | -.12 | .01 | **.43** | **-.28** | **-.36** | -.11 | -.19 | **-.50** | **.19** | **-.25** | **-.70** |  | **-.92** |
| Old | .03 | .06 | **-.35** | .18 | **.20** | .18 | .17 | **.41** | -.06 | **.30** | **.59** | **-.92** |  |

Note: Bold face indicates significant correlation, p < .05.

*Voting Regressed against Measured Variables.* Betas for the regression of voting on each of the 13 demographic variables alone are shown in Table 3, Column 3. Ten of the 13 factors are significantly related to voting. Renters, Minority, Low Income, Service Job, Unemployed and Female were positively related to bond support. Married, High Income, Home

Value and Management Job were negatively related. Betas for College, Parent and Old were small and insignificant.

A multiple regression analysis using all 13 predictors was performed and the resulting betas are shown in the fourth column of Table 3. Of the 10 demographic factors which had significant simple betas, only 3 had significant beta weights in the multiple regression analysis: Married and High Home Value whose effects were negative, and Female, which had a positive effect. In addition, College which had an insignificant simple regression beta, had the second greatest beta weight in the multiple regression analysis. This inconsistency indicates that College is a suppressor variable (Krus and Wilkinson 1986).

Relative information weights for each demographic variable are shown in the last column of Table 3. These weights assess the independent importance of factors separate from the other factors which might be included in the regression analysis. Married again has the highest value on this measure, but it is closely followed by Renter, Minority, and High Home Value all of which have could have important effects on voting.

**TABLE 3**

**Proportion of Positive Referendum Votes Regressed on Demographic Measures**

| Measures | Previous Research | Beta Simple Regression | Beta Multiple Regression | Relative Information |
|---|---|---|---|---|
| Married | ~ | **-.542** | **-.604** | 13.40 |
| Renter | + | **.515** | .012 | 11.10 |
| Minority | + | **.478** | .189 | 10.80 |
| High Income | + | **-.467** | .247 | 8.50 |
| Home Value | ~ | **-.446** | **-.321** | 12.50 |
| Low Income | + | **.420** | .004 | 7.20 |
| Service Job | ~ | **.381** | .046 | 6.90 |
| Management Job | + | **-.336** | .076 | 6.00 |
| Unemployed | ~ | **.308** | -.044 | 3.90 |
| Female | ~ | **.245** | **.194** | 8 |
| College | + | -.064 | **.490** | 7.90 |
| Parent | + | .034 | .089 | 2.20 |
| Old | - | -.002 | -.203 | 1.70 |

Notes: Bold face indicates significant betas, p < .05. Multiple $R^2$ = .45 (adj = .40); significant, p < .001.

*Principal Component Regression.* Principal components were extracted from the correlation matrix of demographic measures, Table 2, and the subsequent space was subjected to a varimax rotation. Results are shown in Table 4. Eigenvalues and percent of variance for the extracted components are shown on the top lines of the table. Over 95 percent of the variance in the original measures is accounted for by the first seven components. Eigenvalues for these seven went from 4.22 down to .82. For the remaining components eigenvalues ranged downward from .18 and percentage of variance averaged .81. Demographic information is well accounted for with seven principal components.

TABLE 4

**Principle Component Analysis and Regression, with Varimax Rotation**

| Components | C 1 | C 2 | C 3 | C 4 | C 5 | C 6 | C 7 | Max values C8 - C13 |
|---|---|---|---|---|---|---|---|---|
| Interpretation | Low SES | Maturity | Home Val | Unemp | Female | Serv Job | College | |
| eigenvalue | 4.22 | 2.58 | 1.52 | 1.22 | 1.01 | .99 | .82 | .18 |
| % of Variance | 32.49 | 19.82 | 11.71 | 9.42 | 7.78 | 7.61 | 6.34 | 1.36 |
| **Factor Loadings** | | | | | | | | |
| Married | -.83 | .06 | .32 | -.27 | -.07 | -.23 | .12 | { |
| Renter | .93 | .02 | -.12 | .18 | .07 | .19 | -.02 | |
| Minority | .64 | -.35 | -.41 | .28 | -.01 | .20 | -.14 | |
| High Income | -.72 | .19 | .46 | -.27 | -.07 | -.20 | .21 | |
| Home Value | -.50 | .21 | .77 | -.24 | -.04 | -.14 | .19 | |
| Low Income | .89 | .15 | -.14 | .15 | .01 | .18 | -.14 | \|.4\| |
| Service Job | .51 | .14 | -.16 | .03 | -.01 | .83 | -.05 | |
| Manage Job | -.48 | .39 | .49 | -.24 | .00 | -.26 | .36 | |
| Unemployed | .39 | -.08 | -.20 | .88 | .07 | .02 | -.12 | |
| Female | .07 | .17 | -.02 | .05 | .98 | -.01 | .00 | |
| College | -.24 | .55 | .28 | -.20 | .00 | -.06 | .72 | |
| Parent | .00 | -.94 | -.16 | .08 | -.10 | -.10 | -.17 | |
| Old | .04 | .97 | .04 | .00 | .14 | .01 | .06 | } |
| **Regression** | | | | | | | | |
| Beta | .44 | -.07 | -.27 | .10 | .21 | .15 | .26 | .12 |
| T | 7.06 | -1.05 | -4.33 | 1.57 | 3.27 | 2.46 | 4.08 | 1.91 |
| p < | .000 | .294 | .000 | .118 | .001 | .015 | .000 | |

Note: Entries in last column are maximum values for components 8 through 13. Multiple $R^2$ = .45 (adj = .40); significant, p < .001

Principal components are defined as weighted sums of values on the original measured variables. Weightings, commonly referred to as "loadings," are shown in Table 4. Principal component 1 has high positive loadings from Renter, Minority, Low Income and Service Job, and high negative loadings from Married, High Income and Home Value. It was given the name "Low SES." The second principal component has a high loading from Old, a high negative loading from Parent and a fairly high loading for College. It is termed "Maturity." Each of the remaining five important components is dominated by a positive loading from a single variable: component 3 by Home Value, component 4 by Unemployed, component 5 by Female, component 6 by Service Job and component 7 by College. Loadings on principal components 8 through 13 do not exceed .4 and are mostly below .3.

Voting proportions were regressed against the 13 principal components with a resulting $R^2$=.45 (adj=.40). These statistics are identical to the regression using the 13 measured variables, as is to be expected since the total variance across the two sets of predictors is identical. Beta values are given in Table 4. None of the lower components, eight to 13, contributed significant betas, indicating that they not only accounted for little variance, they had little relationship to voting. Of the top seven components Low SES, Female, Service Job and College had significant positive betas. Home Value had a significant negative beta, and Maturity and Unemployed were not significantly related to voting.

## DISCUSSION

*Ecological Fallacy.* Voters in precincts with more Low SES people, with more people with College experience, with more Females with more people with Service Jobs gave more support to the bond. *One* explanation for these findings is that *individuals* who were Low SES, who had college experience, who were Female and who had Service Jobs were all more likely to vote "yes." Mindful of Robinson's fallacy warning this interpretation is accepted here. Alternative interpretations would depend on causal factors that are not known and not readily apparent. In similar circumstances other researchers have directly confirmed the validity of the individual explanation for aggregated data (Liu 2007; Moorman 1979).

*Measured Demographic Variables*. Relationships between measured demographic variables and bond support varied depending on the regression model used. This inconsistency is to be expected from measures with high multicollinearity as indicated in Table 2. With high inter-correlations between the independent variables, the computed regression effects for a measure will depend on which other measures are included in the model. The very high negative correlation between being a Renter and being Married illustrates the problem. Precincts with more Married and fewer Renters did not support the bond. Is this a negative effect of Married or a positive effect of Renter or a combination of both effects? The correlation between the two variables precludes answering this question. In any set of demographic measures multicollinearity can be expected, and so regression using such measures is bound to produce results difficult to interpret.

*Principal Components*. The principal component analysis offers more clarity. Seven principal components account for almost all the variance in the 13 measured demographic variables. These components cannot be directly observed, but it is possible to infer coherent meanings from the extent of their relationship with the measured variables. The first principal component has high positive loadings on Renter, Minority, and Low Income and high negative loadings on High Income and Married, and thereby earned the name "Low SES." That Married should be negatively related to Low SES may be surprising but is actually consistent with recent demographic trends. Blacks as opposed to Whites, people with high school as opposed to college education and those with lower versus higher incomes are all less likely to be married (Cohn et al. 2011; Halle 2002; Stevenson and Wolfers 2007). This Low SES component accounts for the most variance in the demographic measurements.

Precincts with higher proportions of Old, very low proportions of Parents (children under 18) and moderately high proportions of people with college experience achieve high scores on the second component, "Maturity." The remaining five principal components are dominated by a single measured variable, suggesting an equivalence between latent component and measured variable. It is important to keep in mind, however, that the derived components are mutually independent and that is not the case with the measured variables.

These seven components are valid descriptions of the present set of voters, but do they also apply to other voting groups, or voters in general? The descriptive, exploratory nature of the analysis included no statistical model for addressing this question. Still, two aspects of the results suggest that they would generalize beyond this study. First, almost all of the demographic variance is accounted for by the seven latent variables, and more than half of it by two variables, Low SES and Maturity. This distribution of variance would seem unlikely if the results were due to chance. Second, the measured variables which define the components Low SES and Maturity form intuitively coherent sets. More than this intuitive rationale is needed, however. The analysis should be replicated with new sets of voters in new situations.

Especially interesting would be replication with differing sets of measured variables. If Low SES and Maturity are basic ways in which individuals vary then it is to be expected that they would become apparent in principal component analyses with other groups of citizens *and* other measurements. The other principal components, those associated here with single measured variables, indicate specific and independent dimensions of variation between people. That is, being female, or having gone to college, for examples, sets a person off in ways unrelated to any other individual characteristic. Replication with other citizen groups could verify the independent existence of these singular components and perhaps identify additional ones. Confirmatory techniques for establishing statistical significance of components and their structure exist (Jackson 1991; Jolliffe 1986) and could be applied in future research.

*Component Effects on Voting.* Five component factors were related to the election outcome. Low SES and Service Job components had positive effects, while high Home Value had a negative effect. This might reflect the difference in the cost-benefit balance between two groups. Libraries provide access to print and computer services that more affluent people can obtain privately but which may be beyond the means of Low SES citizens and those with lower paying Service Jobs. Thus, Low SES precincts would place a higher value on libraries. On the other hand, the tax cost is related to Home Value, so people whose Home Values are higher would tend to vote against the bond. Consistent with this pattern are previous findings that Renters, Minority voters and Low Income voters tend to support bond issues (Cain et al. 1997; Priest and Fox 2005 and Zimmer et al. 2011). *Inconsistent* with this pattern, however, are findings that High Income voters also support bonds (Tosun, Williamson and Yakolev 2012).

Citizens with College experience supported the bond. This would be expected if those who go to college place a higher value on the text-based services available from libraries. Plutzer and Berkman also found that voters who had attended college provided more support for bonds (Plutzer and Berkman 2005). Finally, the component Female was related to bond support. This finding should probably be taken at face value: women value libraries a bit more than men.

Some demographic factors were not related to bond support in the present election. In the past, districts with a high proportion of Parents voted for bonds (Baldson and Brunner 2003) and those with a high proportion of Old voted against (Bowers and Lee 2013). With the citizens in this study the two measures, Parent and Old, had a high negative correlation which helped define a latent component, Maturity. The component Maturity was *not* related to bond support. In this way the present results depart from previous findings.

*Generalization*. Bonds are proposed for many different types of projects: schools, government buildings, sewer systems, libraries and so on. It would be expected that the types of people who would support, or oppose, a bond would vary with the nature of the project. The demographic components which influenced this election for a library may not be important in a school bond election. That said, it is noted that some of the relationships found in earlier school bond elections also held in this library election, namely Low SES and College tended to vote for both types of bonds. Perhaps there are at least some traits which dispose a person to be supportive, or not supportive, of government projects in general.

Hopefully, the demographic components defined by the principal component analysis in this study can be generalized and will describe different electorates. The potential number of direct demographic measures for a population is very large and likely to be highly inter-correlated. Relating such measures to voting is bound to result in the same type of inconsistencies found with direct measures in this election. With the independent factors derived from principal component analyses this confusion is avoided. As stated above replication to verify and/or modify the demographic structure found here would be very useful.